\documentclass[12pt]{iopart}
\usepackage{graphicx,iopams}
\def\pmb#1{\setbox0=\hbox{#1}
\kern-.025em\copy0\kern-\wd0
\kern.05em\copy0\kern-\wd0
\kern-.025em\raise.0433em\box0}

\newcommand{\text}[1]{\rm #1}

\begin{document}

\title{Fixed points in frustrated  magnets revisited}

\author{B. Delamotte$^{1}$, Yu. Holovatch$^{2,3}$, D. Ivaneyko$^{4}$, D. Mouhanna$^{1}$
 and M. Tissier$^{1}$}

\address{$^{1}$ LPTMC, CNRS-UMR 7600, Universit\'e Pierre
et Marie Curie, 75252 Paris C\'edex 05, France}

\address{$^{2}$
Institute  for Condensed   Matter  Physics,  National  Acad.  Sci.   of
Ukraine, UA--79011 Lviv, Ukraine}

\address{$^{3}$
Institut   f\"ur  Theoretische  Physik, Johannes Kepler  Universit\"at
Linz, A-4040 Linz, Austria}

\address{$^{4}$
Ivan Franko National University of Lviv, UA--79005 Lviv, Ukraine}

\begin{abstract}

We analyze the validity of  perturbative renormalization group  estimates
obtained within the   fixed dimension approach of frustrated  magnets.
We reconsider  the resummed  five-loop $\beta$-functions obtained within
the  minimal subtraction scheme  without $\varepsilon$-expansion for both
frustrated magnets and the  well-controlled ferromagnetic systems with
a  cubic  anisotropy.   Analyzing the convergence   properties  of the
critical exponents in these two  cases we find   that the fixed  point
supposed to  control the second order   phase transition of frustrated
magnets is  very likely an unphysical one.   This is  supported  by its
non-Gaussian character at  the  upper critical  dimension  $d=4$.  Our
work confirms  the weak first  order  nature of  the phase  transition
occuring at three dimensions  and provides elements towards a  unified
picture of all existing theoretical approaches to frustrated magnets.

\end{abstract}

\maketitle


\vskip 1cm

\section{Introduction.}

  Although undoubtedly successful to describe the critical behavior of
  $O(N)$-like models, {\it  perturbative}  field theory is still unable
  to  provide a clear, non-controversial  understanding of the physics
  of certain more complex models among which are the famous Heisenberg
  or  $XY$  frustrated magnets (see \cite{delamotte03}  and references
  therein).   At the core  of the problem is  that  different kinds of
  perturbative approaches, performed up   to five- or  six-loop order,
  lead to contradictory results: in dimension $d=3$, first order phase
  transitions    are    predicted  within      the  $\varepsilon$     (or
  pseudo-$\varepsilon$)- expansion
\cite{antonenko95,holovatch04,calabrese03c} whereas a second order
transition  is   found    in the   fixed-dimension  (FD)  perturbative
approaches   performed        either   in    the   minimal-subtraction
($\overline{\hbox{MS}}$) scheme {\it  without} $\varepsilon$-expansion
\cite{calabrese04} or in the massive scheme \cite{pelissetto01a}. 
 In fact,  FD results for frustrated magnets  are neither supported by
 experiments nor by Monte Carlo simulations
\cite{delamotte03,itakura01,peles04,bekhechi06,quirion06} (see however
\cite{calabrese04} where a scaling behavior is found).  They
also disagree with the results  obtained from the non-perturbative
renormalization group (NPRG)  approach    \cite{delamotte03,tissier00}
that predicts  (weak)  first  order  phase  transitions  in  $d=3$  in
agreement with the $\varepsilon$-expansion analysis.

In this article  we shed  light  on the discrepancies encountered   in
perturbative approaches to frustrated  magnets by showing that  the FD
approaches lead to dubious predictions  as for the critical physics in
$d=3$.  Our  key-point relies on  the very nature of the FD
perturbative approach  and is easy to  grasp  already for the simplest
---  $O(N)$    ---    model.   In  this   case,   the   (non-resummed)
renormalization group   (RG)  $\beta$-function  at   $L$ loops   is  a
polynomial in  its  coupling constant $u$   of order $L+1$.  Thus,  it
admits $L+1$  roots  $u^*$, $\beta(u^*)=0$,  that  are either real  or
complex.  Within the $\varepsilon$-expansion, when one solves
the    fixed  point   (FP)   equation  in   successive orders   in
$\varepsilon=4-d$, the only   non-trivial  FP retained  is  by
definition such that   $u^*\sim \varepsilon$.   On  the contrary,  in the    FD
approaches, when one directly (analytically or numerically) solves the
non-linear FP equation at fixed $d$ (fixed  $\varepsilon$) no real root
can be a priori discarded. As a result, the  generic situation is that
the number of FPs as well as their stability  vary with the order $L$:
at a given order, there can exist several  real and stable FPs or none
instead of a single  one. This artefact of the  FD approach is already
known   and  was   first noticed   in  the  massive   scheme  in $d=3$
\cite{parisi80}.  The way to cope with it is also known: resumming the
perturbative expansion of  $\beta(u)$ (see  e.g.  \cite{zinnjustin89})
is  supposed both to restore  the  non-trivial Wilson-Fisher FP and to
suppress the  non-physical or ``spurious'' roots.   This is indeed what
occurs for the $O(N)$ model for which the FP analysis performed on the
resummed  $\beta$-function of  FD  approaches enables  to discriminate
between  physical and ``spurious''  FPs.  However this  ability of the
resummation  procedures to   eradicate spurious  solutions  of the  FD
approach has never  been questioned and,  {\it a fortiori}, evaluated
in  the context of more complex  models and, in particular, for models
with several coupling constants.  We precisely argue, in this article,
that  the  situation is very   different  for frustrated  magnets  and
probably   for  several   other   models.   Indeed,   considering  the
$\beta$-functions derived at  five loops in the $\overline{\hbox{MS}}$
scheme and using a standard resummation procedure
\cite{calabrese04} we show that the FP found in $d=3$ without
expanding in $\varepsilon$, although it persists after resummation, is in
fact  spurious.  

 Our conclusion is  based on   several facts: (i) the
critical exponents computed at the  FP supposed to control the  second
order   behavior of    frustrated  magnets  display  bad   convergence
properties with the order of computation in the controversial cases of
XY and  Heisenberg spin systems (ii) when  analyzed with the  same FD
approach, the  field theory relevant    to ferromagnetic systems  with
cubic    anisotropy displays a  similar  FP  --- having no counterpart
within the $\varepsilon$-expansion  ---  in contradiction with  its  well
established critical physics.  The critical exponents computed at this
supernumerary FP display the same bad  convergence properties with the
loop order as in the case of XY and  Heisenberg frustrated magnets  (iii) the coordinates
$(u_1^*,u_2^*)$ of the attractive  FP  found in the FD approach of frustrated magnets
are multivalued functions of $(d,N)$ --- $N$ being the number of spin
components --- because  of the existence  of a topological singularity
in     the   mapping between    $(d,N)$   and    the   FP  coordinates
$(u_1^*(d,N),u_2^*(d,N))$.    This    singularity   provides    strong
indications  of  the existence  of   pathologies in the  RG  equations
obtained at fixed dimension  (iv) finally, we provide strong arguments
showing that the supernumerary  FPs found in  the frustrated and cubic
models survive in  the upper critical  dimension $d=4$ where  they are
found   to  be non-Gaussian,  a  behavior   deeply connected  with the
existence of the above mentioned topological singularity.  Being given
the present state of knowledge of $\phi^4$-like theories that are very
likely trivial in $d=4$, this fact confirms  the serious doubts on the
actual existence of these supernumerary FPs.

 \section{Resummation method.}

 To  investigate  the five-loop   $\beta$  functions  derived in   the
 $\overline{\hbox{MS}}$  scheme   we have to    resum  them.    Before
 discussing  the case of a series  of  two coupling constants relevant
 for frustrated magnets or  ferromagnetic models with cubic anisotropy
 we recall, for the sake of clarity, the main steps  needed to resum a
 series of  one  coupling constant    $u$ as well  as  the  underlying
 hypotheses  (see for a   review \cite{suslov05}). Let  us consider  a
 series
\begin{equation}
f(u)=\sum_{n} a_n \ u^n \
\label{series1}
\end{equation}
where the coefficients $a_n$ are supposed to grow as $n!$.

The Borel-Leroy sum associated with $f(u)$ is given by:
\begin{equation}
B(u)=\sum_{n} {a_n\over \Gamma[n+b+1]} \ u^n \
\label{borelsum}
\end{equation}
and is supposed to converge, in the complex  plane, inside a circle of
radius $1/a$, where  $u=-1/a$ is the  singularity of $B(u)$ closest to
the  origin.   Then,        using  this   definition  as    well    as
$\Gamma[n+b+1]=\int_0^{\infty} t^{n+b}\ e^{-t} dt$, one can rewrite
\begin{equation}
f(u)= \sum_{n} {a_n\over \Gamma[n+b+1]}  \ u^n \int_0^{\infty} \  dt \
e^{-t}\ t^{n+b}
\end{equation}
and, interchanging summation and integration, one can {\it  define} the
Borel transform of $f$ as:
\begin{equation}
f_B(u)=\int_0^{\infty} \ dt \ e^{-t}\ t^{b}\ \ B(ut)\ .  \
\label{boreltrans}
\end{equation}

In order  to perform the integral in  (\ref{boreltrans})  on the whole
real positive  semi-axis one has to  find  an analytic continuation of
$B(t)$.    Several methods  can   be   used, Pad\'e   approximants for
instance.   However,   it is generally   believed  that the   use of a
conformal  mapping  is more  efficient   since  it makes  use  of  the
convergence  properties of the Borel  sum.   Under the assumption that
all the singularities of $B(u)$ lie on the negative real axis and that
the Borel-Leroy sum is analytic in  the whole complex plane except for
the cut extending from $-1/a$ to $-\infty$, one can perform the change
of variable:
\begin{equation}
\omega(u)={\sqrt{1 + a\, u}-1\over \sqrt{1 + a\, u}+1}
 \hspace{1cm}  \Longleftrightarrow     \hspace{1cm}  u(\omega)={4\over
 a}{\omega\over(1-\omega)^2}
\label{conformal}
\end{equation}
that maps the  complex $u$-plane cut from  $u=-1/a$  to $-\infty$ onto
the unit circle in the $w$-plane such that the singularities of $B(u)$
lying  on the negative axis   now lie on   the boundary of the  circle
$|w|=1$.  The   resulting expression $B(u(\omega))$ has   a convergent
Taylor  expansion  within the  unit circle   $|\omega|<1$  and can  be
rewritten:
\begin{equation}
B(u(\omega))=\sum_{n} d_n(a,b) \hspace{0.1cm} \left[\omega(u)\right]^n
\label{borel3}
\end{equation}
where    the coefficients $d_n(a,b)$  are      computed so  that   the
re-expansion  of   the r.h.s.  of    (\ref{borel3})  in powers of  $u$
coincides  with  that   of   (\ref{series1}).   One  obtains   through
(\ref{borel3}) an analytic  continuation  of $B(u)$ in  the  whole $u$
cut-plane  so that  a resummed  expression of  the  series  $f$ can be
written:
\begin{equation}
f_R(u)=\sum_{n} d_n(a, b) \hspace{-0.1cm} \int_0^{\infty}
\hspace{-0.2cm}dt\, \, {e^{-t}\, t^{b}\ \left[\omega(u t)\right]^n}\ . 
\label{resummation1}
\end{equation}
        
In practice it  is   interesting  to   generalize  the
expression  (\ref{resummation1}) by   introducing \cite{kazakov79} the
expression

\begin{equation}
f_R(u)=\sum_{n} d_n(\alpha,a, b) \hspace{-0.1cm} \int_0^{\infty}
\hspace{-0.2cm}dt\, \, {e^{-t}\,  t^{b}}\ { \left[\omega(u t)\right]^n \over 
\left[1-\omega(u t)\right]^{\alpha} }\  
\label{resummation2}
\end{equation}
whose meaning  will be explained  just below.

If  an infinite  number of terms   of the series  $f_R(u)$ were known,
expression (\ref{resummation2}) would be independent of the parameters
$a$ and $b$ and $\alpha$. However  when only a  finite number of terms
are known, $f_R(u)$  acquires a dependence  on them. In principle, the
parameters $a$  and $b$ are  fixed by the  large order behavior of the
series:
\begin{equation}
a_{n\to\infty}\sim (-a_0)^n \, n!\, n^{b_0}
\end{equation}
which  leads  to $a=a_0$   and  $b\simeq b_0+3/2$  \cite{leguillou80}   while
$\alpha$ is determined by the  strong coupling behavior of the initial
series:
\begin{equation}
f(u\to\infty) \sim u^{\alpha_0/2} \
\end{equation}
which can  be  imposed  at any  order of   the  expansion  by choosing
$\alpha=\alpha_0$.  The common assumption is that  the above choice of
$a$, $b$ and  $\alpha$   improve the convergence  of   the resummation
procedure since it encodes exact results.

 Let us however emphasize that, often, only  $a$ is known and that the
 other parameters, $\alpha$ and $b$, are considered either as free (as
 for instance in  \cite{calabrese04}) or variational (for  instance in
 \cite{mudrov98c} where  $\alpha$   is determined by    optimizing the
 apparent convergence of the series). In  any case,  the choice of value
 of  $a$, $\alpha$ and $b$ must  be validated a posteriori by checking
 that a small change of  their value does  not yield strong variations
 of   the  quantities under study.     Such  variations would  clearly
 indicate that  one    has chosen  an   unstable,  parameter-dependent
 calculation  procedure   or  that  one has   not    computed  the
 quantities under study at a  sufficiently high order of  perturbation
 theory  to consider them  as converged.   In the  following, we shall
 employ  this ``stability criterion'' to validate  -- or invalidate --
 the  results obtained by  means  of the  FD perturbative  approach to
 frustrated magnets.

In   the  context  of   frustrated magnets,  the above described  resummation
procedure must be  extended to  several  (two) coupling constants.  In
this  case, $f$ is a function  of two variables  $u_1$ and $u_2$ known
through its  series  expansion in  powers   of $u_1$  and $u_2$.   The
resummation  technique   used    in   \cite{calabrese04}  consists  in
considering $f$ as a function of $u_1$ and $z=u_2/u_1$:
\begin{equation}
f(u_1,z)=\sum_{n} a_n(z) \ u_1^n \
\label{series}
\end{equation}
and in resumming with respect to  a single variable $u_1$  only.  An important hypothesis
underlying this procedure is  that one can safely resum (\ref{series})
with respect to  $u_1$  while keeping  $z$  fixed, {\it i.e.}  without
resumming with respect to   $u_2$.  Under this hypothesis  the   resummed
expression associated with $f$ reads:
\begin{equation}
f_R(u_1,z)=\sum_{n} d_n(\alpha,a(z),b;z) \hspace{-0.1cm}
\int_0^{\infty}
\hspace{-0.2cm}dt\, \, {e^{-t}\, t^{b}}{ \left[\omega(u_1 t;z)\right]^n \over 
\left[1-\omega(u_1 t;z)\right]^{\alpha} }
\label{resummation}
\end{equation}
with:
\begin{equation}
\omega(u;z)={\sqrt{1 + a(z)\, u}-1\over \sqrt{1 + a(z)\, u}+1}\ 
\end{equation}
where,  as     above,  the  coefficients     $d_n(\alpha,a(z),b,z)$ in
(\ref{resummation})  are computed  so that   the  re-expansion of  the
r.h.s.  of (\ref{resummation}) in powers of $u$ coincides with that of
(\ref{series}).

\section{Frustrated magnets.}  

 The Hamiltonian  relevant for frustrated
 systems is given by:
\begin{equation}
\begin{array}{ll}
\displaystyle \hspace{0cm}{\mathcal H}= \int{\rm d^d} x \Big\{\frac{1}{2}
\left[(\partial\pmb{$\phi$}_1)^2+
 (\partial\pmb{$\phi$}_2)^2                     +                  m^2
 (\pmb{$\phi$}_1^2+\pmb{$\phi$}_2^2)\right]+\\
\\
\hspace{2.3cm}\displaystyle \frac{u_1}{4!}\ \left[\pmb{$\phi$}_1^2+
\pmb{$\phi$}_2^2\right]^2 
+\frac{u_2}{12}\ \left[(\pmb{$\phi$}_1 \cdot \pmb{$\phi$}_2)^2-
\pmb{$\phi$}_1^2\,\pmb{$\phi$}_2^2\right] \Big \}
\end{array}
\label{landau}
\end{equation}
 where the  $\pmb{$\phi$}_i$, $i=1,2$, are $N$-component vector fields
 and  $u_1$ and $u_2$ are  the coupling constants that satisfy $u_1>0$
 and $u_2<4 u_1$ --- which  corresponds to an Hamiltonian bounded from
 below. For $m^2>0$ the ground state  of Hamiltonian (\ref{landau}) is
 given  by $\pmb{$\phi$}_1=\pmb{$\phi$}_2=0$ while  for  $m^2<0$ it is
 given by a configuration  where $\pmb{$\phi$}_1$ and $\pmb{$\phi$}_2$
 are orthogonal with   the same norm.   The Hamiltonian (\ref{landau})
 thus describes a symmetry breaking scheme between a disordered and an
 ordered phase where  the $O(N)$ group  of rotation is  broken down to
 $O(N-2)$ which is relevant for frustrated magnets (for details see
\cite{delamotte03} for instance).

Let us  first recall  the  FP structure of  Hamiltonian (\ref{landau})
{\it around $d=4$} at leading order in $\varepsilon=4-d$
\cite{garel76,bailin77,yosefin85}.  For $N$ larger than a critical
value  $N_c(d)$ depending on the dimension,  the RG equations display,
apart the usual Gaussian  ($u_1^*=u_2^*=0$) and $O(2N)$ ($u_1^*\ne  0,
u_2^*=0$)  FPs, two non-trivial ($u_1^*\ne  0$ and $u_2^*\ne 0$) FPs :
one,  $C_+$, is stable;  the  other  one,  $C_-$,  is unstable.  Above
$N_c(d)$, the transition is thus predicted to be  of second order.  As
$N$ is  lowered starting from values  of $N>N_c(d)$, the two FPs $C_+$
and $C_-$ get  closer  and finally  collapse together for  $N=N_c(d)$.
Below $N_c(d)$, there is no  longer a stable  FP and the transition is
expected to  be of first order.  The  value of $N_c(d)$  for $d=3$ has
been computed by several means: in a  double expansion in $u_1$, $u_2$
and $\varepsilon=4-d$ up to five-loops
\cite{antonenko95,holovatch04,calabrese03c}, directly in $d=3$ in a 
weak-coupling expansion within the massive scheme up to six-loops
\cite{pelissetto01a}, within a NPRG approach \cite{delamotte03} and, finally, 
within        the  $\overline{\hbox{MS}}$      scheme {\it    without}
$\varepsilon$-expansion
\cite{calabrese04}.  The predictions obtained within the perturbative approaches performed 
at fixed $d$ strongly differ from  those obtained using other methods,
in particular  in  $d\le 3$, see below.  This  is the  reason which   led us to
reconsider this kind of approach.

We thus   apply the resummation     procedure described above    without
$\varepsilon$-expansion \cite{schloms87}  to      the    $\beta_{u_i}$
functions,    $i=1,2$,    obtained   at      five   loops    in    the
$\overline{\hbox{MS}}$ scheme
\cite{calabrese04}. More precisely, as in \cite{calabrese04}, we resum
$(\beta_{u_i}(u_1,z) +  \varepsilon u_i)/{u_1}^2$, $i=1,2$, instead of
$\beta_{u_i}(u_1,z)$, which, in  fact, leads  to similar results.  For
the model (\ref{landau}), the  region of Borel-summability is given by
$2u_1-u_2\ge 0$ (see for instance
\cite{calabrese04} for details) to which corresponds the value $a(z)=1/2$. 
 For $2 u_1-u_2\le 0$ there exists a  singularity on the positive real
 axis so  that the  series is no  longer Borel-summable.   However, as
 noted   in \cite{calabrese04},   as   far as  $4u_1-u_2\ge   0$,  the
 singularity of the Borel transform closest to  the origin is still on
 the negative axis.  Thus, the asymptotic  behavior is still correctly
 taken  into  account by the conformal   mapping and one  can,  {\it a
 priori},  trust the  resummed  results.   Note  finally that $b$  and
 $\alpha$ are typically varied in the ranges $[6,30]$ and $[-0.5,2]$.

\begin{figure}[htbp] 
\vspace{0cm}
\hspace{1cm}
\includegraphics[width=0.8\linewidth,origin=tl]{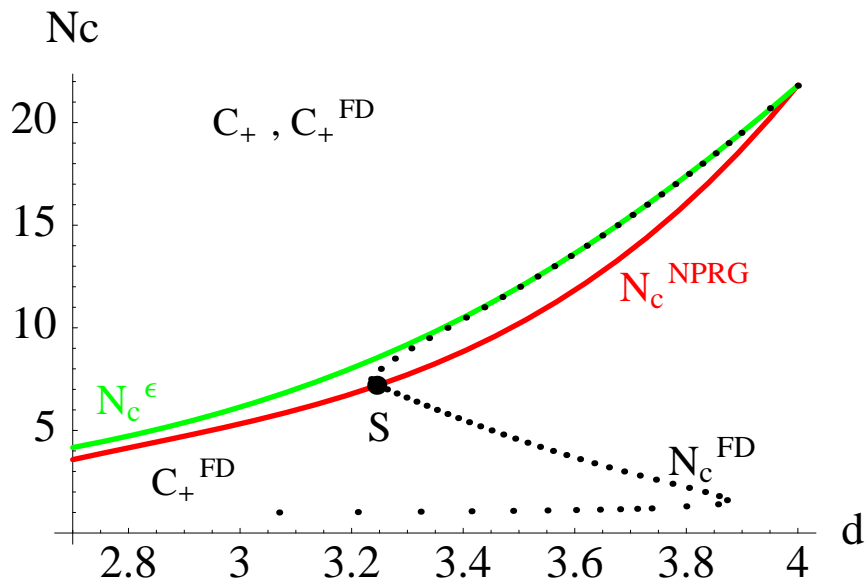}\hfill%
\caption{Curves $N_c(d)$ obtained within the $\overline{\hbox{MS}}$
  scheme       with    $\varepsilon$-expansion ($N_c^\varepsilon$),  without
  $\varepsilon$-expansion  ($N_c^{\rm   FD}$)   and  the  NPRG   approach
  ($N_c^{\text{NPRG}}$).   The  resummation     parameters    for  the
  $\overline{\hbox{MS}}$ curve are $a=1/2$, $b=10$ and $\alpha=1$.}
\label{courbes_ncd} 
\end{figure}

 One finds, in agreement with
\cite{calabrese04}, that there exists a curve (parametrized by
$N_c(d)$ or its reciprocal $d_c(N)$) such that for $d<d_c(N)$ a stable
FP $C_+$  governs the critical properties of  the system.   The curves
$N_c(d)$ obtained within this scheme is shown in Fig.\ref{courbes_ncd}
by a  dotted line ($N_c^{\rm FD}$).  On  the same  figure we  show the
results for $N_c(d)$ obtained within the NPRG approach
\cite{delamotte03}, red solid curve ($N_c^{\text{NPRG}}$)  and by (resummed)
 $\varepsilon^5$-expansion
\cite{calabrese04}, green solid curve ($N_c^\varepsilon$).  Two points 
must be  noted.    First     the  curves  $N_c^\varepsilon(d)$     and
$N_c^{\text{NPRG}}(d)$   show  a  remarkable
agreement   being given the     very  different nature   of the    two
corresponding computations.  Second, one can see on Fig.\ref{courbes_ncd} the strong
discrepancy  between the two  previous approaches and the perturbative
FD  approach.  In particular the S-like   shape of the curve $N_c^{\rm
FD}(d)$ obtained  within  the perturbative  FD  approach is  such that
the FP $C_+^{\rm FD}$ exists for all $N\geq 2$ in $d=3$ contrary to the other
approaches in which a FP $C_+$ exists only for $N>N_c(d=3)\simeq5$
\cite{delamotte03}. 

 This situation, put together with the fact, already underlined in the
 Introduction, that the FD  approach  a priori displays  spurious FPs,
 lead us to strongly question the validity  of the results obtained at
 FD and,  in  particular, the existence  of  genuine FPs in  $d=3$ for
 $N=2,3$.

\section{Convergence of the loop expansion.}

 \subsection{The frustrated  model.}  We first examine the convergence
 of the loop expansion of the  FD approach by studying the sensitivity
 of   the  resummed quantities  with  respect   to variations   of the
 resummation parameters $b$  and $\alpha$ as well as  the order $L$ of
 the computation.

  We  focus on the  (real part of the)  correction to scaling exponent
  $\omega$  at  the   FP $C^+$   which rules    its  stability  :  for
  Re($\omega)>0$ the  FP is stable  and  for  Re($\omega)<0$ it is unstable. 
   Within the  FD approach,  one finds  in $d=3$  for $N<7$
  that $C^+$  is a focus, that is  $\omega$ is complex  at  this FP (the
  flow spirals around it). In practice, following \cite{mudrov98c}, we
  optimize $\omega(\alpha,b,L)$ by first  choosing $\alpha$  such that
  $\omega(L+1)-\omega(L)$  is  minimal.    Then, one   determines  the
  parameter $b$ in such a way that $\omega(b)$ is stationary.  We have
  checked that similar   convergence properties are obtained for   the
  exponent $\nu$ \cite{delamottenext}.

\begin{figure}[htbp] 
\vspace{-0cm}
\hspace{2cm}
\includegraphics[width=0.7\linewidth,origin=tl]{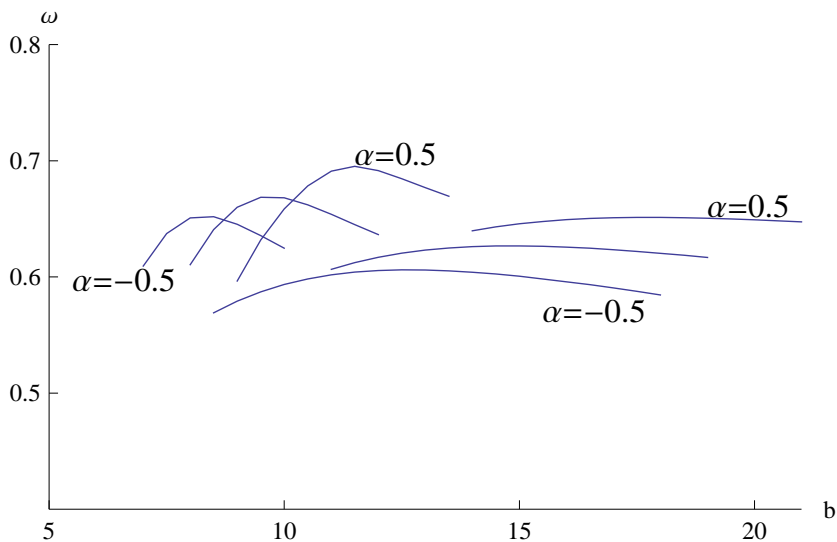}\hfill%
\caption{The critical exponent $\omega$ for $N=9$ as a function of $b$
 at four (curves on the left) and five (curves on the right) loops for
 $\alpha=-0.5,\, 0,\, 0.5$ for the frustrated model.}
\label{frustreN9} 
\end{figure}

 We start by  considering   the uncontroversial case of   a  ``large''
 number of components of the order parameter, typically greater than 7
 in   $d=3$.  Indeed,   in this  case,  {\it   all}  perturbative  and
 non-perturbative approaches  agree together  and  find a  stable FP 
 characterizing a second order  behavior \cite{delamotte03}.  In
 Fig.\ref{frustreN9},  we  display $\omega(b)$ in   the case $N=9$ for
 $L=4$ and  $L=5$ for typical  values of $\alpha$.   At a given order,
 one sees that it is  indeed possible to find values  of $b$ that make
 $\omega$ stationary.  By performing  the same  analysis at four-  and
 five-loop orders  one observes (i)   that the dependence of  $\omega$
 with  respect to $b$  decreases with  the order  of  the expansion as
 expected  (ii) a  convergence  of  the results with  the
 order with,  however,  large  error  bars typically  around
 $5-7\%$. Note that the typical error  bars  obtained  for the Ising model
 with the same methodology at the same orders  are much less than
 1$\%$.

\begin{figure}[htbp] 
\vspace{-0cm}
\hspace{2cm}
\includegraphics[width=0.7\linewidth,origin=tl]{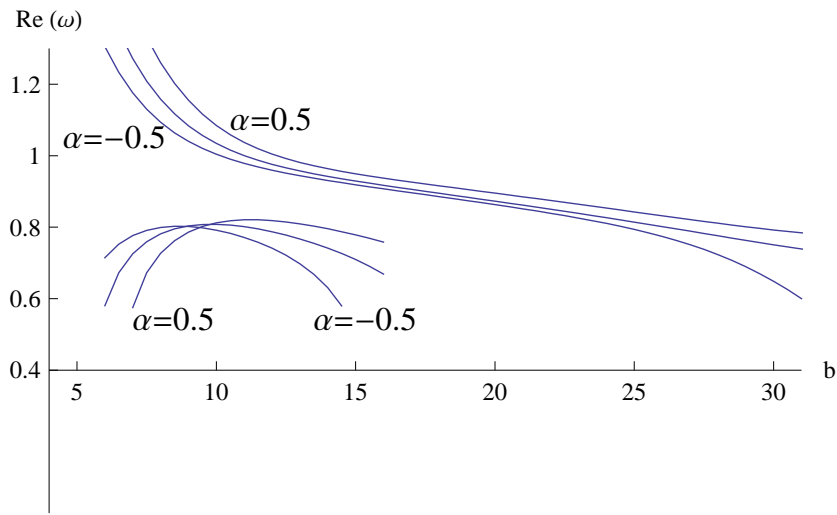}\hfill%
\caption{The (real part of the) critical exponent $\omega$  for $N=3$ as a
 function of $b$
 at  five  (upper  curves)    and four   (lower    curves) loops   for
 $\alpha=-0.5,\, 0,\, 0.5$ for the  frustrated model. Note the change of
 scale with respect to Fig.\ref{frustreN9}.}
\label{frustreN3} 
\end{figure}

 We then consider the controversial  case  of Heisenberg systems ($N=3$).   In
 Fig.\ref{frustreN3}   we display again $\omega(b)$ for $L=4$ and $L=5$.
 There, a  new phenomenon occurs:  while one still  finds a stationary
 value of $\omega(b)$  at four loops,  this is  no longer  the case at
 five loops.    Moreover, considering the  values  of $\alpha$ and $b$
 that minimize the difference $\omega(L=5)-\omega(L=4)$ one observes a
 bad ``convergence'': the errors bar on the critical exponents are now
 of order $30\%$ and thus far larger than in  the case $N=9$.  An even
 worse behavior is obtained in the XY case.

From   these  analyses one  gets   striking indications of drastically
different convergence properties for the $N=9$ and  $N=3$ cases.  This
suggests a qualitative difference between  the corresponding FPs.  To characterize more
precisely this  difference let us now study  the cubic model along the
same lines.

 \subsection{The cubic model.}

 We  now consider the ferromagnetic model  with cubic anisotropy whose
 Hamiltonian is:
\begin{equation}
  \displaystyle \hspace{0cm}{\mathcal H}= \int{\rm d^d} x
\Big\{\frac{1}{2}\left[(\partial\pmb{$\phi$})^2+ m^2
\pmb{$\phi$}^2\right]+ {u\over 4!}  \left[\pmb{$\phi$}^2\right]^2
+{v\over 4!}\sum_{i=1}^N \phi_i^4\Big \}
\label{cubic}
\end{equation}
with  a  $N$-component vector field $\pmb{$\phi$}$.   The Hamiltonian
(\ref{cubic})  is  used  to  study the critical   behavior of numerous
magnetic  and  ferroelectric systems  with appropriate order parameter
symmetry (see e.g.  \cite{folk00b}).   The $\beta$-functions are known
at    five-loop   order    in the     $\overline{\hbox{MS}}$    scheme
\cite{kleinert95}  and  at six-loop    order  in the   massive  scheme
\cite{carmona00}. Apart from the  Gaussian ($u^*=v^*=0$) and an  Ising
FP ($u^*=0,  v^*\ne 0)$, there exist  two FPs more: the $O(N)$ symmetric FP
$(u^*\neq0,  v^*=0)$ and the mixed one   $M$ $(u^*\neq 0, v^*\neq 0)$.
The Ising and  Gaussian FPs are both  unstable for all values  of $N$.
The  $O(N)$  FP is stable   and  $M$ is  unstable  with  $ v^*< 0$ for
$N<\tilde{N_c}$ and the opposite   for $N>\tilde{N_c}$.  The  critical
value $\tilde{N_c}$ has been found  to  be slightly  less than 3:  for
instance $\tilde{N_c}\sim 2.89(4)$ in
\cite{carmona00} and $\tilde{N_c}\sim 2.862(5)$ in \cite{folk00b}.

Let us now analyze the FP structure  of the model (\ref{cubic}) within
the  $\overline{\hbox{MS}}$ scheme  without $\varepsilon$-expansion by
applying the conformal mapping  Borel transform (\ref{resummation}) at
$d=3$   ($\varepsilon=1$).     The parameter  $a(z=v/u)$   entering in
(\ref{resummation}) is now    given   by $a(z)=1+z$ for    $z>0$   and
$a(z)=1+z/N$ for $z<0$ while the region  of Borel-summability is given
by the    condition $u+v>0$ and   $Nu+v>0$.  Within  this   scheme one
surprisingly  observes that, in addition  to the above mentioned usual
FPs, there exist,  in a whole domain  of parameters  $b$ and $\alpha$,
several   other     FPs  that   have     no    counterparts   in     the
$\varepsilon$-expansion.  In particular, one of  them that we call $P$
(which is stable and such that $u^*>0, v^*<0$) exists for any value of
$N\lesssim 7.5$ and lies  in the region of  Borel-summability $u+v>0$.
The   presence of this FP,  if   taken seriously, would have important
physical  consequences  since it would correspond   to  a second order
phase transition    with a new universality   class.   However no such
transition has  ever been reported.   On the  contrary, a first  order
behaviour for  all values of   $N$ larger than  $\tilde{N_c}$ is found
within perturbative
\cite{carmona00,calabrese03d} or non-perturbative \cite{tissier01b}
field theoretical analysis as well as numerical simulations
\cite{itakura99} in related systems (four-state antiferromagnetic
Potts model). Thus, the existence  of $P$ has to  be considered as  an
artefact of  the FD analysis.   Note finally, and  interestingly, that
$P$ is found to  be a {\it  focus} FP,  a striking similarity with  the
frustrated case.

\begin{figure}[htbp] 
\vspace{-0cm}
\hspace{2cm}
\includegraphics[width=0.7\linewidth,origin=tl]{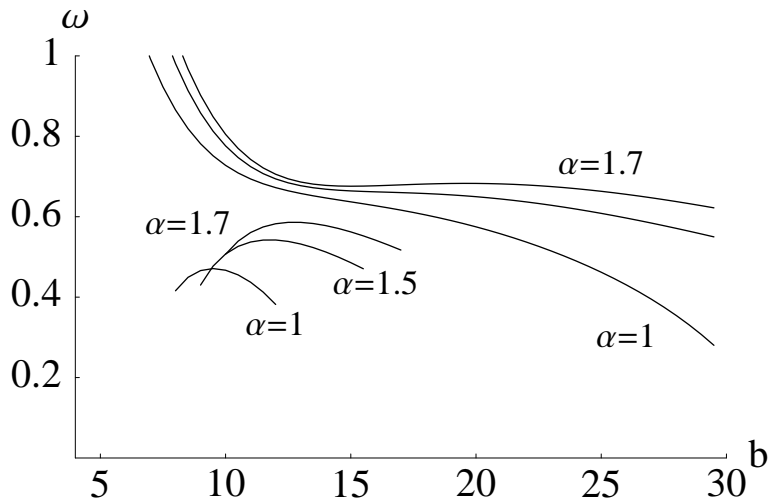}\hfill%
\caption{The critical exponent $\omega$ as a function 
 of $b$ at five (upper curves)
 and four (lower  curves) loops  for  $\alpha=1,1.5,1.7$  for   the
 cubic model (N=2).}
\label{omegacubic} 
\end{figure} 

At this stage, it is very instructive to  perform the same convergence
analysis as the one performed for $C_+$ in  the frustrated case. The Ising, $O(N)$ 
and mixed  $M$
FPs display  good  convergence properties and we
focus in the following on the supernumerary  FP  $P$. In Fig.\ref{omegacubic}
we plot   $\omega(b)$ at this  FP for $L=4,5$  and  for three different
values of $\alpha$.  Interestingly, when comparing Fig.\ref{frustreN3}
and Fig.\ref{omegacubic} we find  similar behavior  between the  cubic
case      and the  frustrated       case  for    $N=3$.   Indeed,   on
Fig.\ref{omegacubic} one finds stationary  values of $\omega(b)$  at
four loops but, at five loops,  this is only the case  for $\alpha=1.7$.
It is also interesting to consider the values of $\alpha$ and $b$ that
minimizes the  difference $\omega(L=5)-\omega(L=4)$.  From there, one  finds error
bars for $\omega$ of order $40\%$, i.e. of the {\it same} order of magnitude as the
one found in the $N=3$ frustrated case.  Being given the fact that the
FP $P$ is clearly an  artefact of the FD  analysis, our study suggests
that the lack of  convergence  of $\omega(b)$ {\it characterizes}  the
behavior at  a spurious FP. Thus,  coming back  on the frustrated case
one is naturally led  to the conclusion  that the FPs $C_+^{\text{FD}}$
found in $d=3$ and $N=2,3$ should also  be interpreted as spurious FPs,
as  artefacts of the FD analysis.  In order to confirm this statement
we  now examine   another   characteristic  feature  of  the  set   of
$C_+^{\text{FD}}$ FPs considered as functions of $d$ and $N$.

\section{The singularity $S$.}

 We now   display a specific  feature of  the   RG flow  of frustrated
 magnets analyzed within the FD approach  that provides another strong
 indication of   the problematic character  of this  approach.  Let us
 consider  the   coordinates  $(u_1^*,  u_2^*)$  of   the FP
 $C_+^{\text{FD}}$ as functions  of  $d$ and $N$:  $u_i^*=u_i^*(d,N)$,
 $i=1,2$.  These  functions are the  roots of the $\beta$-functions of
 the couplings  $u_1$  and  $u_2$ obtained  in   the  FD approach  and
 resummed   according    to      the   scheme     sketched      above,
 Eq.(\ref{resummation}):
\begin{equation}
\beta_{u_1}(u_1^*,u_2^*)=\beta_{u_2}(u_1^*,u_2^*)=0 \ . 
\end{equation}
 The  resummed $\beta$-functions are smooth functions  of  $d$ and $N$
 showing   no  particular     feature    for  $2<d\le4$    and   $2\le
 N<\infty$. However, as we now show, the functions $u_i^*=u_i^*(d,N)$,
 $i=1,2$ exhibit  a  non-trivial behavior as  $d$  and $N$ are  varied
 continuously around the point labelled  $S$ in Fig.\ref{courbes_ncd}.

\begin{figure}[htbp] 
\vspace{-0cm}
\hspace{2cm}
\includegraphics[width=0.7\linewidth,origin=tl]{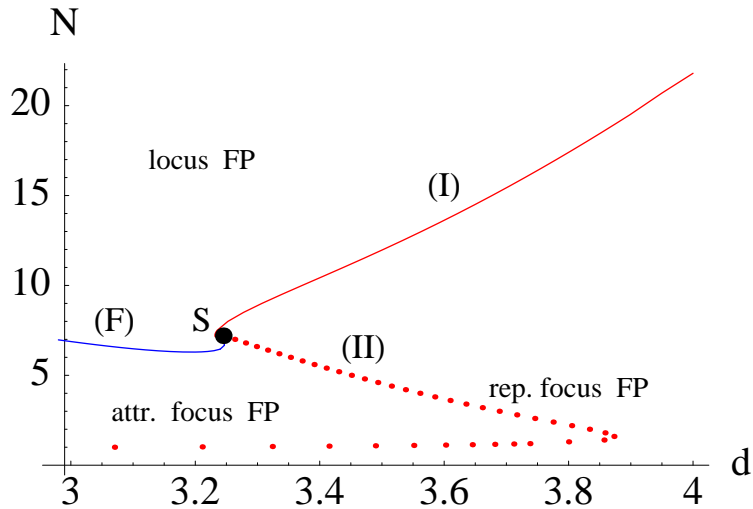}\hfill%
\caption{Curve $N_c(d)$  within the $\overline{\hbox{MS}}$
  scheme  without $\varepsilon$-expansion.  The   part   (I) of  the   curve
  $N_c^{\text{FD}}$ corresponds to  a boundary between a region where,
  at a given dimension $d$ there exists a stable locus FP at large $N$
  and no FP at small $N$. The part (II) of the curve $N_c^{\text{FD}}$
  corresponds to a boundary between a region where  at fixed $N$ there
  exists an attractive focus FP  for $d<d(N)$ and a repulsive focus  for
  $d>d(N)$.  Finally the line (F) is defined as follow: above  (F), $C_+^{\text{FD}}$ is a  locus FP
  ($\omega_1$ and $\omega_2$ are both real)  and below (F), it is
  a focus ($\omega_1$ and $\omega_2=\omega_1^*$ are complex).  }
\label{focuslocus} 
\end{figure}

 Let us first give, in Fig.\ref{focuslocus}, the precise definition of
 the curve $N_c(d)$  in  the FD approach.  This  curve is made  of two
 parts: the first  one, labelled (I), corresponds to   the part of  the
 curve above $S$ while the second one, labelled  (II), corresponds to the
 part below $S$.  (I)  is  defined as the  line  in the $(d,N)$  plane
 above which   there exist two  non-trivial {\it   locus} FP  (that is
 having real exponents $\omega$), one stable $C_+^{\text{FD}}$ and one
 unstable $C_-^{\text{FD}}$, and below which   there exists none.   We
 recall  the mechanism of   disappearance of these  FPs: when  {\it at
 fixed   dimension $d$}, $N$ is  decreased  from large  values down to
 $N=N_c(d)$  the two  FPs $C_+^{\text{FD}}$ and  $C_-^{\text{FD}}$ get
 closer  and closer  from each other   and  finally collapse right  on
 (I).  For  $N$ below $N_c(d)$, the  coordinates  $u_1^*$ and $ u_2^*$
 become  complex and there is no  longer any non-trivial FP with real
 coordinates. Thus, the part (I)  of the curve $N_c(d)$ corresponds to
 the region where  the speed of the  RG flow between $C_+^{\text{FD}}$
 and  $C_-^{\text{FD}}$ vanishes. The  same   behavior is observed  in
 other ($\varepsilon$-expansion and NPRG) approaches and the numerical
 values  of $N_c(d)$ in the corresponding  part  of the curve are very
 close     when         calculated    by        different    approches
 (c.f. Fig.\ref{courbes_ncd}).

As  usual, the  speed  of the  RG flow around   a  FP  can be
obtained by linearizing  the flow equations at  this point. It is thus
governed by the two eigenvalues of the ``stability'' matrix
\begin{equation}
M_{ij}={\frac{\partial\beta_{u_i}(u_1,u_2)}{\partial
u_j}}{\Bigg\vert_{u_1^*, u_2^*}}
\end{equation}
that provide the  speeds of the RG  flow in its two eigendirections at
the   FP  considered.  These  eigenvalues   define  the  two  critical
exponents  $\omega_1$  and $\omega_2$   governing  the  corrections to
scaling. The equation of the part (I) of $N_c(d)$ is thus given by:
\begin{equation}
\omega_1=\omega_1\bigg(u_1^*(d,N), u_2^*(d,N)\bigg)=0
\end{equation}
where   $\omega_1$ is  the  eigenvalue  of   $M$  corresponding to the
eigendirection    of   the    flow     joining   $C_+^{\text{FD}}$  to
$C_-^{\text{FD}}$.  Now,  when  moving   {\it on} the   line  $N_c(d)$
towards  $S$  one  finds   that  $\omega_2$ decreases  and  eventually
vanishes. One thus defines the point $S$ by:
\begin{equation}
\omega_2(u_1^*(S), u_2^*(S))=0\ . 
\end{equation}
Thus,  right  at $S$,  $\omega_1=\omega_2=0$  and, with the  choice of
parameters  $a=1/2, b=10, \alpha=1$ one   finds that at $S$, $d=3.24$,
$N=7$ and $(u_1^*(S), u_2^*(S))=(0.3,0.7)$.
Thus,  $S$ is just  outside  --- but   not  far of ---  the  region of
Borel-summability.   Note  also    that, since  its coordinates   verify $4u_1^*-
u_2^*>0$, $S$  is  still in the  region where  the resummed results  are
trustable.

   In fact, the point $S$ is also a special point in the sense that it
   is   the   endpoint   of      another line,  labelled        $(F)$ on
   Fig.\ref{focuslocus}  which is such   that  below $(F)$, the  two
   exponents $\omega_i$'s acquire  an imaginary  part and are  complex
   conjugates: $\omega_1=\omega_2^*$.  This  means that,  below $(F)$,
   the FPs become focuses, either  attractive or repulsive.  This is  in
   particular the  case  of  the    FP  $C_+^{\text{FD}}$ found     in
   \cite{calabrese04} in $d=3$.  From now on  we concentrate on the FP
   $C_+^{\text{FD}}$    since  the coordinates  of   $C_-^{\text{FD}}$
   rapidly  grow and go outside  the region of Borel summability. Note
   that the part of (F) shown on  Fig.\ref{focuslocus} lies inside the
   region of  Borel summability and thus, within  the  FD approach, is
   supposed to be well under control.

 We now  define  the  part (II) of  the  curve  $N_c(d)$ as   the line
 separating the region  where $C_+^{\text{FD}}$ is an {\it attractive}
 focus FP  (for   $d<d_c(N)$) and the    region  where it  is a   {\it 
 repulsive}   focus  FP  (for  $d>d_c(N)$)  \footnote{  The change  of
 stability of $C_+^{\text{FD}}$ occurs as  follows. For $d<d_c(N)$ and
 $N<7.5$  and sufficiently  close to (II),  $C_+^{\text{FD}}$ is a  focus
 and is attractive inside a basin of attraction which is a limit cycle
 for  the RG flow.  This  limit cycle is repulsive  : inside it the RG
 flow  converges to $C_+^{\text{FD}}$, outside  the  RG flow diverges.
 When, at fixed  $N$, the dimension $d$  is increased this limit cycle
 shrinks and becomes a point on  (II).  When $d$ is further increased,
 the limit  cycle   re-appears    but  now  it  is    attractive   and
 $C_+^{\text{FD}}$  becomes   a   repulsive  focus FP.}.      Thus, by
 definition,  (II)   is the  line   on  which the    real part of  the
 $\omega_i$'s vanishes:
\begin{equation}
\text{Re}(\omega_1)=\text{Re}(\omega_2)=0\ 
\end{equation}
but it is {\it no longer}  a line  where $C_+^{\text{FD}}$ collapses with another
FP and  disappears.   It is remarkable  that  within the  FD approach,
there exists a line (the  part (II)  of the  curve $N_c(d)$) that  has
{\it no counterpart} in other approches.

Contrary  to what occurs above, the coordinates $u_1^*$ and $u_2^*$
of $C_+^{\text{FD}}$ on the  line (II) are, for  a large part of (II),
far outside the region of Borel summability.   It is thus not possible
to compute accurately  the  location of (II). We,   however, emphasize
that only the existence  of (II) is  necessary for the validity of our
arguments, not its precise determination. As for the existence of this
part (II)  of the curve $N_c(d)$, it  is an unavoidable consequence of
the existence of $S$  which is 
supposed to be under  control {\sl  within the perturbative  FD  approach}.
Thus, either $S$ really  exists and part (II) of $N_c(d)$ also exists --- with
a shape that   could   be somewhat  different of   the  one  drawn  on
Fig.\ref{focuslocus} --- or it does not exist and neither does $S$. In
this second case, this would mean that the {\it whole}  resummation scheme is
questionable at least   for sufficiently low  $d$  and $N$  (typically
$d<3.2$   and $N<7$).   We argue  in   the following  that this is  very
probably what occurs.

\begin{figure}[htbp] 
\vspace{-0cm}
\hspace{1cm}
\includegraphics[width=0.7\linewidth,origin=tl]{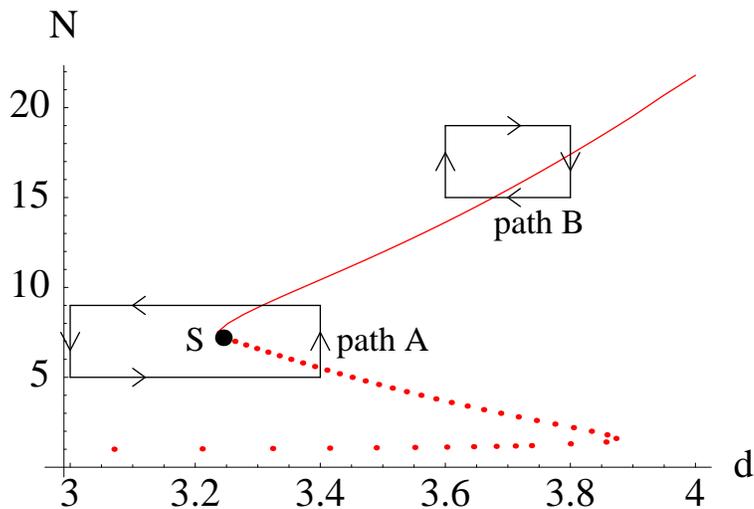}\hfill%
\caption{Two different paths in the $(N,d)$ plane. Path A encloses
 $S$ and is such  that the coordinates  of the FP $C_+^{\text{FD}}$ do
 not go back to their original values after a  trip along this path at
 variance with path B.  }
\label{paths} 
\end{figure}

Let us now show on Fig.\ref{paths} that there exists a very non-trivial property of the
RG flow which is a consequence of the  existence of $S$.   The idea is to
follow  continuously     the   coordinates   of   the   FP
$C_+^{\text{FD}}$   along    a  path  encircling  $S$,      path A  in
Fig.\ref{paths} for instance. We start, for instance at $(d=3, N=5)$
go to $(d=3.4, N=5)$ then to  $(d=3.4, N=9)$ then  to $(d=3, N=9)$ and
finally go back to $(d=3, N=5)$. The surprising fact is that by making
a trip  along such a closed  path, the coordinates $(u_1^*, u_2^*)$ of
$C_+^{\text{FD}}$  do not go  back to  their  original value. This  is
specific  to $S$ since  along any  closed path that  does not encircle
this  point ---   path B in  Fig.\ref{paths}   for instance ---  the
coordinates $(u_1^*,  u_2^*)$ of $C_+^{\text{FD}}$   always go back to
their original values. Let us emphasize here that such a path can well
cross part (I) of the curve $N_c(d)$ as  path B does in Fig.\ref{paths}.  In this
case,  $C_+^{\text{FD}}$  has complex coordinates  on  the part of the
path  which is below   (I),  but as the   path  crosses again  (I) the
coordinates  become real again and go   back finally to their original
values.  All this makes $S$ a topological singularity of the functions
$u_1^*(d,N)$ and $ u_2^*(d,N)$.

 Let us remark at this stage  that, before having drawn any conclusion
 from  the  existence of a  topological  singularity, one faces a very
 unusual  property  of the mapping  from the  plane $(N,d)$ to  the FP
 coupling constants space $(u_1^*, u_2^*)$ which makes multivalued the
 functions $u_1^*(N,d)$  and $u_2^*(N,d)$.  While  one cannot a priori
 discard  the    possibility  of such a   behavior    of the functions
 $u_1^*(N,d)$ and $u_2^*(N,d)$, it is  tempting to attribute it to the
 coexistence at a  given $(d,N)$ of  FPs  that are identical to  those
 found within the $\varepsilon=4-d$  expansion and FPs that are artefacts
 of the FD  approach.  Let us now  draw  the full consequences  of the
 striking behavior described above.

\section{Fixed point at the upper critical dimension.}

 We now  present our last argument in  favor of the spurious character
 of the FPs obtained within the FD approach for small $d$ and $N$.  It
 is based, to a large extent, on the  existence of the singularity $S$
 in the $(u_1^*,u_2^*)$ plane that leads to a striking property of the
 field theory describing frustrated magnets.

Let us first  recall some basic features about  the description of the
long   distance  physics of a  lattice   system by a  continuous field
theory.  The most  important  ingredient in  this construction is  the
choice  of  a low-energy effective  Hamiltonian in  terms of the order
parameter   $\phi$.  This choice  implies  the selection   of a finite
number of terms ---  the $\phi^2$ and $\phi^4$  terms for second order
phase transitions --- among  the infinite number of operators obtained
in  the    Hubbard-Stratonovitch    derivation of     the  microscopic
hamiltonian.  This selection of  the most relevant terms fully  relies
on the existence  of an upper critical  dimension where  the theory is
{\it  perturbatively} infrared free,  {\it   i.e.}  controlled by  the
Gaussian    FP.   Indeed, it  is     only  under this  condition  that
power-counting  makes  sense  since it  is  based  on  the engineering
dimension  of the   field  that neglects fluctuations.    Perturbation
theory can then   be used  since,  by definition,  it consists  in  an
expansion around the Gaussian  theory.  Also, the perturbative results
obtained in this way are reliable as long as the (infrared attractive)
FP found this way (i) is connected to  the Gaussian by the RG
flow, (ii) lies in  the Borel-summability region (iii)  is not too far
from the Gaussian FP so that  calculations performed at $L$ loops lead
to  converged  results.  Let  us indicate that    because of point (i)
above, {\it if}  the theory is trivial  in $d=4$, it is  very probable
that   any  non-trivial   FP identified    in  $d=3$, once   followed
continuously from $d=3$ to $d=4$,  becomes Gaussian in this dimension.
This is  in particular  what underlies the validity of the $\varepsilon=4-d$ expansion.

Let   us now  briefly  discuss the   question of  triviality of scalar
theories in $d=4$ and the  ensuing consequences for their perturbative
analysis  in any dimension below 4.   We first emphasize that there is
no   rigorous proof  of   the   triviality   of  scalar  theories   in
$d=4$. However (i) there is a large body of evidences of triviality at
least for the $O(N)$ models and in particular for the Ising model (ii)
it is very likely  that even if the  scalar models were non-trivial in
$d=4$,   perturbation   theories  would  not  be   able   to reach  the
corresponding FP  \cite{callaway88}. Thus, the most natural hypothesis
is that the $O(N)\times O(2)$ theory is also trivial in $d=4$. An ever
weaker hypothesis, that we make and use in the following,  is that this is the
case when it  is analyzed perturbatively.  All previous considerations
and assumptions lead us to  conclude that, very probably, any physical
FP found {\it perturbatively} in  any dimension must be Gaussian  when
followed continuously in  $d=4$. Let us  notice that a non-perturbative
approach, performed by  some of the present authors \cite{delamotte03}
on the $O(N)\times O(2)$ theory did   not  lead to any  non-trivial   FP in  $d=4$ which  sustains our
triviality hypotheses in $d=4$.

Therefore,  according  to our   hypotheses, a  practical way  to check
whether a FP  found at  a given  dimension,  $d=3$ for instance, is  a
genuine FP or is just an artefact of perturbation  theory is to follow
it by continuity up to $d=4$
\cite{holovatch04,dudka04}.   If the FP  survives as a non-Gaussian FP
at this  dimension  we consider it as   spurious.  It is important  to
realize  that our   criterion does not    exclude  FP in  $d=3$  whose
coordinates  become complex between  $d=3$   to $d=4$ as  far as  they
vanish in $d=4$. This  is in particular the  case  for all FPs associated to the 
paths  with fixed $N$ in the $(d,N)$ plane that cross (I) above $S$.

\begin{figure}[htbp] 
\begin{center}
\includegraphics[width=0.6\linewidth]{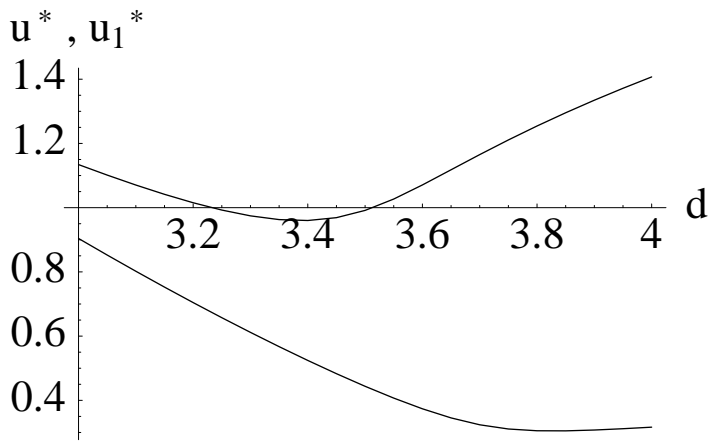}
\end{center}
\caption{The $u^*$ coordinate of the FP $P$  of the cubic model ($N=2$, upper curve) and the 
$u_1^*$ coordinate of the FP $C_+^{\text{FD}}$ of the frustrated model ($N=3$, lower curve) as
functions of $d$.}
\label{frustre3} 
\end{figure}

Let  us  apply this  criterion to  analyze  the  FPs  $P$ and
$C_+^{\text{FD}}$ that appear in the FD analysis of the cubic and frustrated models. 
 We present our results in Fig.\ref{frustre3} where
we  have displayed the  coordinates $u^*$  and $u_1^*$ associated with
these FPs  as functions of $d$ at   fixed $N$.  Manifestly,  they both
survive everywhere above $d=3$ and are {\it not} Gaussian in $d=4$. In
the cubic   case, this is  true  for all values of   $N$ for  which $P$
exists.  In the frustrated case, this is true for $N$ typically below 7. 
According  to  our criterion, $P$ is,  as
expected, always  found  to  be  spurious while $C_+^{\text{FD}}$   is
spurious only for $N<7$.  We  thus  conclude, in the frustrated  case,
that the FPs found in $d=3$ for $N=2,3$ are spurious.

Let us emphasize that when $d$ approaches 4 the coordinates of the FPs
$P$  and $C_+^{\text{FD}}$ for $N<7$  become  large and  do no  longer
belong  to the  region  of Borel-summability so  that   they cannot be
determined  accurately as  it  is the case of  part  (II) of the curve
$N_c(d)$.  One could thus naively conclude  that it is not possible to
safely decide whether these FPs are or are not Gaussian in $d=4$. This
is actually not the case  for at least two  reasons. First, if these
FPs {\it  were} Gaussian in $d=4$, their  coordinates  just below this
dimension would be extremely small and their existence as Gaussian FPs
in $d=4$ could be safely  established within perturbation theory  even
without any resummation procedure. Since for $d$  just below 4 no such
FPs  close to the Gaussian  are  found in  perturbation theory,  their
non-Gaussian   character  is doubtless  in   $d=4$.    Second, in  the
frustrated case, the  non-Gaussian character of the  $C_+^{\text{FD}}$
FP  is a clear  consequence of  the existence  of (II) which,  itself,
relies  on the existence of  the singularity $S$.  Indeed, following a
FP $C_+^{\text{FD}}$ along  a path starting  in $d=3$, going  to $d=4$
and crossing $N_c^{\text{FD}}(d)$  {\it above} $S$, the coordinates of
$C_+^{\text{FD}}$  become complex in $d=d_c(N)$ and  both the real and
the imaginary parts of $u^*_1$ and $u^*_2$ go to  zero for $d=4$ where
it is   thus a Gaussian   FP.  If, on  the contrary,  the path crosses
$N_c^{\text{FD}}(d)$ {\it below} $S$, $C_+^{\text{FD}}$ changes from a
stable focus to an unstable  one at $d=d_c(N)$ and  does not go to the
Gaussian in $d=4$.  The  singularity $S$ is therefore  responsible for
the non-Gaussian character of  $C_+^{\text{FD}}$ in $d=4$. Thus,  even
if the coordinates of $C_+^{\text{FD}}$ for $N<7$ become large for $d$
close  to 4, its  non-Gaussian character  in  this dimension  makes no
doubt.   Note finally    that it is  tempting   to follow this  FP
$C_+^{\text{FD}}$ {\it above} $d=4$ where there exist rigorous proofs
of  the  triviality of  the        scalar $\phi^4$  field       theory
\cite{aizenmann81}. Indeed,  as it is suggested by Fig.\ref{frustre3},
the FP $C_+^{\text{FD}}$ apparently  survives at finite distance above
d=4. However this last fact must be taken with great caution since 
$C_+^{\text{FD}}$  is now  deeply  out of the   region of Borel-summability and  one has  no longer  
any control of where the FP really  lies.

\section{Conclusion.}

\begin{figure}[htbp] 
\vspace{0cm}
\hspace{1cm}
\includegraphics[width=0.8\linewidth,origin=tl]{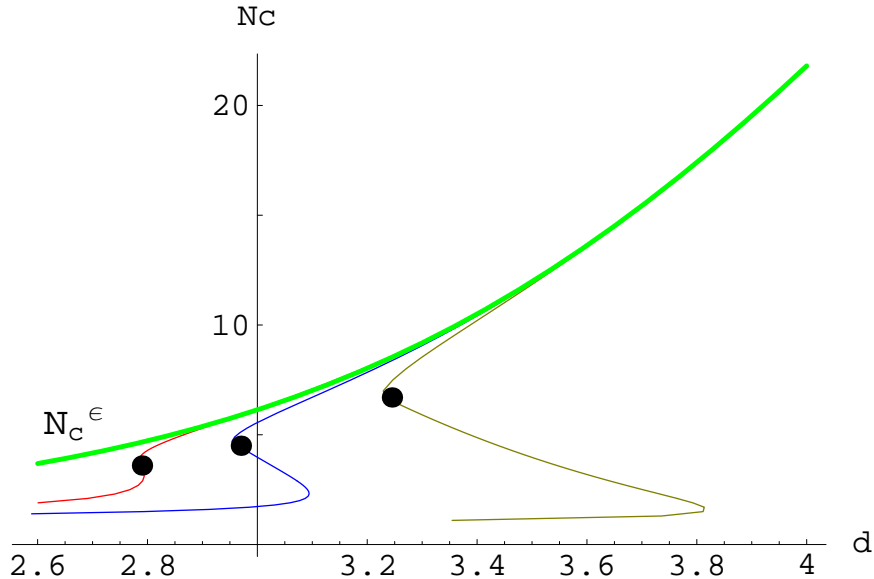}\hfill%
\caption{Three curves $N_c^{\text{FD}}(d)$ for different values of the 
 parameter $a$ (from right to left $a=0.5, 1.3,  1.5$)  and the  curve $N_c^{\varepsilon}(d)$. 
  The  parameters $b$  and $\alpha$ are  $b=10$ and
 $\alpha=1$.  The parts of the curves  below the black dots correspond
 to a regime of Borel non-summability. }
\label{nc} 
\end{figure}

    It   appears   from  our  study  that     the  FPs
$C_+^{\text{FD}}$   identified  in  the FD  approach   are  very likely
spurious.  The transition in frustrated  magnets should thus be of ---
possibly  weak   ---   first    order in  agreement   with   NPRG  and
$\varepsilon$-expansion approaches.  It remains to explain the failure in
the    resummation     procedure   used    in   the      FD  approach,
Eq.(\ref{resummation}).  As  already  emphasized this procedure relies
on  the hypothesis that  resumming with  respect to  $u_1$, keeping  a
polynomial   structure in  $u_2$,   is sufficient.   Alternatively,  a
resummation  of the  series with respect   to  the {\it two}  coupling
constants could   be  required to obtain   reliable  results  (see for
instance   \cite{alvarez00}     for   the randomly     diluted   Ising
model). Postponing these considerations for a future publication
\cite{delamottenext} we assume that the use of Eq.(\ref{resummation})
as such is justified.   Then, a possible origin  of the failure in the
resummation   procedure could be  that  the  series considered are not
known at large enough order to reach the asymptotic behavior.  In this
case there   would  be no reason  to  fix  the parameter $a$    at its
asymptotic   value  $a=1/2$ and one  should  have to vary   it as $b$ and
$\alpha$  to optimize   the results \cite{mudrov98c}.   We  display in
Fig.\ref{nc} the  curves $N_c^{\text{FD}}(d)$  for different values of
$a$.  The part corresponding  to large values  of $N_c$, typically for
$N_c\gtrsim 7$, is almost insensitive to the variations of $a$ whereas
this  is clearly  not  the  case  for smaller   values of   $N_c$.  In
particular, for  sufficiently    large  values   of    $a$,  typically
$a\geq1.5$,  the  S-like   part    is  pushed below    $d=3$  so  that
$N_c^\varepsilon(d)$, $N_c^{\text{FD}}(d)$ and $N_c^{\text{NPRG}}(d)$ are
then compatible  everywhere for  $3<d<4$.  Let us  also notice in this
respect that, for  $a=1.3$,  the shape  of the curve $N_c^{\text{FD}}(d)$
is even compatible with the results obtained in the massive scheme in $d=3$
in which one finds FPs for all values of $N$ but in the range
$5.7(3)<N<6.4(4)$ \cite{calabrese03b}.   This suggests that the two FD
methods  ---  massive  scheme  in  $d=3$ and  $\overline{\hbox{MS}}$  without
$\varepsilon$-expansion ---  are in fact compatible  but  suffer from the
same problems of convergence. Thus, under our  hypothesis,
all qualitative differences between the different approaches disappear,
so that  the problem   would boil  down  to a   question of  order  of
computation.

Finally  note that our present considerations surely pertains to the case of
frustrated magnets in $d=2$ \cite{calabrese02}. Indeed we have checked
that  the  FP found    in $d=2$ is   continuously  related  to  the FP
$C_+^{\text{FD}}$  in $d=3$ which  makes  its existence doubtful.  Our
conclusions could also  apply in other situations  where FPs that have
no  counterpart in  the  $\varepsilon$-expansion  are  found,  as it  is the
case, for instance, in QCD at finite temperature \cite{basile04}.

 We wish to thank P. Azaria, P. Calabrese, R. Folk, R. Guida and J. Zinn-Justin
 for useful discussions. Work of Yu.H. was supported in part by the  Austrian Fonds zur 
F\"orderung der wissenschaftlichen Forschung, Project P19583. We acknowledge 
the CNRS-NAS Franco-Ukrainian bilateral exchange program. 

\section*{References.}

\providecommand{\newblock}{}

\end{document}